# Cyclization of Short DNA Fragments

Pui-Man Lam[*]
Physics Department, Southern University
Baton Rouge, Louisiana 70813

and

Yi Zhen[+]
Department of Natural Sciences, Southern University
6400 Press Drive
New Orleans, Louisiana 70126

*puiman_lam@subr.edu;  +yzhen@suno.edu

**Abstract:** From the per unit length free energy for DNA under tension, we have calculated an effective contour length dependent persistence length for short DNA. This effective persistence length results from the enhanced fluctuations in short DNA. It decreases for shorter DNA, making shorter DNA more flexible. The results of the J-factor calculated using this effective persistence length are in good agreement with experimental data.

## I. Introduction

DNA molecules are semi-flexible polymers whose mechanical properties are well-described by the worm like chain (WLC) model [1], using a single fitting parameter, the persistence length $l_p = 50$ nm, or about 150 base pairs (bps). The WLC model reproduces the experimental force-extension curve with impressive precision, for DNA with contour length $L$ much larger than the persistence length [2]. However, for short DNA fragments, with contour length less than or equal to the persistence length, new experimental results [3,4,5] suggest a length dependent flexibility, with effective persistence lengths much shorter than that of the WLC model.

In order to describe a contour length dependent persistence length for short DNA, Xu et al [6] introduce a correlated worm like chain model (CWLC). In the WLC model, the persistence length $l_p$ describes the correlation between the unit tangent vectors $\hat{t}(s_1)$ and $\hat{t}(s_2)$ at two contour positions $s_1$ and $s_2$ along the chain: $<\hat{t}(s_1)\cdot\hat{t}(s_2)> \sim e^{-|s_1-s_2|/l_p}$, with $|s_1 - s_2| \gg l_p$. In the CWLC model an extra length scale $l_D \sim 10$ bps is introduced, that describes the additional short-ranged correlation between the derivatives of the tangent vectors $\left\langle \frac{d\hat{t}}{ds}(s_1) \cdot \frac{d\hat{t}}{ds}(s_2) \right\rangle \sim e^{-|s_1-s_2|/l_D}$, which is absent from the WLC model. In the CWLC model, a contour length $L$ dependent, effective persistence length $l_{EP}(L)$ can be obtained that depends explicitly on the parameters $l_D$ and the original persistence length $l_p$. Moreover, this effective persistence length $e^{-|s_1-s_2|/l_p}$ goes over into the original persistence length rapidly when $L >> l_p$.

Xu et al apply this contour dependent effective persistence length in the study of the J-factor, defined as the ratio of equilibrium constants for cyclization and bimolecular association [7] and is a measure of the probability of DNA forming a ring. This J-factor had been studied theoretically by Shimada and Yamakawa in their seminal paper [8]. They obtained the result

$$J_{SY}(L) = \frac{C_1}{L_p^3}\left(\frac{l_p}{L}\right)^6 \exp\left[-\frac{2p^2 l_p}{L} + C_2 \frac{L}{l_p}\right] J_t\left(\frac{L}{l_t}, \frac{L}{h}\right) \quad (1)$$

Here, $C_1$ and $C_2$ are known constants and factor $J_t$ is the contribution of the torsional constraints giving rise to the rapid oscillations of $J_{SY}(L)$ with the period of the helix repeat $h$=10.5 bp. Using the constant value $l_p = 53$ nm, Eqn. (1) gives good agreement with experimental values for large values of $L$. However, for small values of $L < l_p$, it gives values orders of magnitude less than the experimental values. By substituting their effective contour length persistence length into Eqn. (1), and by adjusting the parameter $l_D$, Xu et al are able to find better agreement with experimental data for small values of $L$.

Sinha et al [9] had studied short DNA under tension. They found that for short DNA, the per unit length free energy is modified and depends on the contour length. It approaches the value given by the WLC model only in the limit of infinite $L$. We will show that from this form of the unit length free energy, a contour length dependent persistence length can be obtained. Using this contour length dependent persistence length in Eqn. (1), the resulting J-factor is also in good agreement with experiment. The advantage of this approach is that the contour length dependent persistence length is independent of any extra parameter and results only from the enhanced fluctuations in the short DNA. In section II we present the calculation of persistence length from the per unit length free energy fort short DNA. In section III we will apply this contour length dependent persistence length to the calculation of the J-factor. Section IV is the conclusion.

**II. Free energy and persistence length of short DNA**

Sinha and Samuel [9] had studied DNA under tension $f$. They found that for short DNA under constant tension, the unit length free energy $g(f)$ is given by

$$\exp(L\tilde{g}(\tilde{f})) = \int_0^1 du \exp\left(-\frac{L[\tilde{w}(u) - \tilde{f}u]}{L_p}\right) \quad (2)$$

where the dimensionless quantities $u$, $\tilde{f}$, $\tilde{g}$ and $\tilde{w}$ are given by

$$u = \frac{X}{L}, \quad \tilde{f} \equiv \frac{fl_p}{k_B T}, \quad \tilde{g}(\tilde{f}) \equiv \frac{g(f)}{k_B T}, \quad \tilde{w}(u) \equiv \frac{L_p W(X)}{Lk_B T}. \quad (3)$$

In the limit of an infinite long chain, $L \to \infty$, the integral can be replaced by the integrand evaluated at the value of $u$ that will give the largest value in the exponent. This is given by the condition

$$\frac{d}{du}[\tilde{w}(u) - \tilde{f}u] = 0 \quad (4)$$

This will yield the usual Legendre transform, where $X$ is the extension,

$$Lg(f) = -W(X) + fX, \quad \text{with} \quad W(X) = \int_0^X f(X')dX'. \quad (4)$$

The force-extension curve in the worm like chain (WLC) model is given by the widely used interpolation formula [1]

$$f = \frac{(k_B T)}{l_p}\left[\frac{X}{L} + \frac{1}{4}\left(1 - \frac{X}{L}\right)^{-2} - \frac{1}{4}\right], \quad (5)$$

Using Eqn. (4), the function $W$ can be calculated analytically:

$$W(X(f)) = L\frac{k_B T}{4l_p}\left[\frac{X_f(f)}{L}\left(2\frac{X_f(f)}{L} - 1\right) + \left(1 - \frac{X_f(f)}{L}\right)^{-1}\right], \quad (6)$$

where $X_f(f)$ is the inverse function of Eqn. (5), which give the extension $X$ as a function of the force $f$. Since the extension is a single-valued, monotonic increasing function of $f$, this can always be done numerically.

Using Eqn. (6) in the form

$$\tilde{w}(u) = u(2u - 1) + (1 - u)^{-1},$$

the unit length free energy for a short DNA from Eqn. (2), is given by

$$\tilde{g}(\tilde{f}) = \frac{1}{L} \ln\left\{ \int_0^1 du \exp\left( -\frac{L}{4l_p}[u(2u-1)+(1-u)^{-1}] + \frac{L}{l_p}\tilde{f}u \right) \right\}. \quad (7)$$

The integral in Eqn. (7) can be easily calculated numerically.
From Eqn. (4), the derivative of the free energy with respective to $f$ is given by

$$Lg'(f) = X_f(f) \quad (8)$$

Therefore knowing the derivative of the free energy with respect to $f$ would give the extension $X_f(f)$ as a function of $f$. This derivative can be obtained from Eqn. (7) as

$$g'(f) = \int_0^1 du\, u \exp\left( -\frac{L}{4l_p}[u(2u-1)+(1-u)^{-1}] + \frac{L}{l_p}\tilde{f}u \right) \left[ \int_0^1 du \exp\left( -\frac{L}{4l_p}[u(2u-1)+(1-u)^{-1}] + \frac{L}{l_p}\tilde{f}u \right) \right]^{-1}$$

(9)

The integrals in Eqn. (9) can be numerically calculated and hence $g'(f)$. Knowing the force-extension curve for the short DNA of contour length $L$, we can determine its effective persistence length $l_{EP}(L)$ by using Eqn. (5) in a least square fit. In Fig. 1 we show the calculated effective persistence length $l_{EP}(L)$ normalized by $l_p$, as a function of the number of base pairs. We have used the value $l_p = 53$ nm, the value used by Xu et al [6]. The effective persistence length decreases as the contour length decreases. This means that shorter DNA is more flexible.

### III. J-factor for short DNA

We have used the effective persistence length $l_{EP}(L)$ calculated in the last section in Eqn. (1) to calculate the J-factor for DNA. The result is shown in Fig. 2. The experimental data of Du et al [10] are shown in open circles. The agreement is reasonable.
However, Vafabakhsh and Ha [11] recently reported a J-factor about three orders of magnitudes larger than the classical results by Du et al [10], at the same contour length of $L = 105$ bps. A comparison of the two experimental methods shows that in the case of ref. [11], there might exist a capture radius, so that the results correspond to the probability for the two DNA end separated by a distance $R$, rather than forming a close ring. For this problem, an analytic approximation within the WLC framework, for this probability $Q_{WLC}(L/l_p, R/L)$, has been provided by Becker et al [12], without taken into account the twist degree of freedom. The twist degree of freedom can be included by multiplying by the factor $J_t\left(\frac{L}{l_t}, \frac{L}{h}\right)$ in Eqn. (1). We have calculated the probability $Q_{WLC}$ by replacing $l_p$ by our effective $l_{EP}(L)$ and the value $R = 11$ nm. The result of our calculation, including the twist degree of freedom is also shown in Fig. 2, compared with the experimental data of ref [11]. The agreement is also reasonable.

## IV. Conclusion

We have calculated the per unit length free energy of short DNA under tension using the theory of Sinha and Samuel [9]. Using this free energy, and within the WLC model, we have calculated an effective persistence length $l_{EP}(L)$, depending on the contour length of the DNA. This effective persistence length results only from the enhanced fluctuations in shorter DNA and decreases as the contour length decreases. This shows that shorter DNA is more flexible. With this effective persistence length we have calculated the J-factor using the Shimada-Yamakawa theory and also the probability that the two DNA ends be separated by a distance $R$. Both quantities are in reasonable agreement with experimental data.

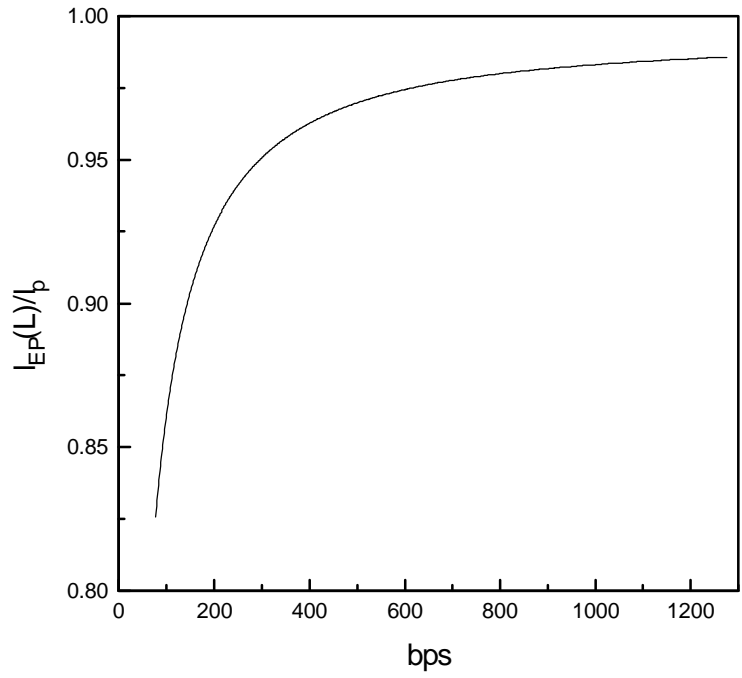

Fig. 1: The effective persistence length $l_{EP}(L)$ normalized by the persistence length $l_p$ as a function of the number of base pairs.

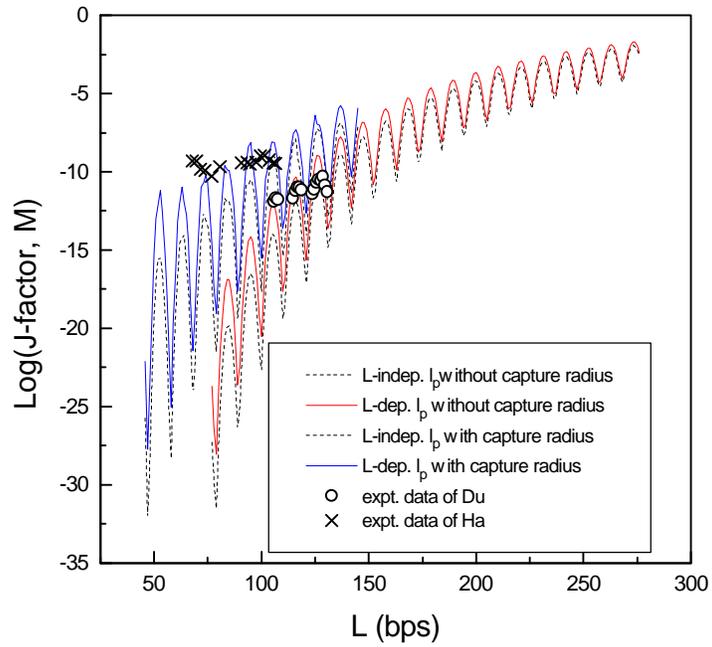

Figure 2. Log (J-factor, M). Dashed lines denote theoretical results obtained with L-independent $l_p$ and solid line denote theoretical results obtained with L-dependent $l_p$. Open circles and crosses denote experimental data of Du without capture radius and Ha with capture radius, respectively.